\documentstyle[prl,aps,psfig]{revtex}
\begin{document}
\draft

\twocolumn[\hsize\textwidth\columnwidth\hsize\csname
@twocolumnfalse\endcsname

\title{Compactness of the set of Faddeev and Lippmann--Schwinger
 equations for the three-body Coulomb problem}
\author{Zolt\'an Papp}
\address{Institute of Nuclear Research of the
Hungarian Academy of Sciences, \\
P.O. Box 51, H--4001 Debrecen, Hungary}
\date{\today}
\maketitle

\begin{abstract}
The set of Faddeev and Lippmann--Schwinger
integral equations for three-body systems involving
Coulomb interactions deduced from a ``three-potential''
picture are shown to be compact for all energies
and a method of solution is given.
\end{abstract}

\vspace{0.5cm}
\pacs{PACS number(s): 21.45.+v, 03.65.Nk, 02.30.Rz, 02.60.Nm}
]

There are two well-known, practical but  genuinely different
 approaches to solve three-body scattering equations of 
Faddeev type and which
involve Coulomb-like interactions.
In the first of these, the Alt--Grassberger--Sandhas (AGS)
equations are solved by the ``screening and renormalizing''
technique. This approach and its applications
are discussed in a recent review~\cite{a-s}.
The alternative approach is 
to solve the Faddeev--Noble differential equations.
To do so, in configuration space, one needs 
asymptotic boundary conditions.
This approach has also been reviewed recently~\cite{f-p}.
Although these approaches have been in use for some time, 
problems still exist with 
the renormalization procedure of method one and with
the approximate boundary condition required in method
two. Consequently, to date, only
limited solutions are available below or above the breakup threshold.

Herein I  propose a new approach for solving the three-body Coulomb
scattering problem.
As usual, one presumes that 
the quantum mechanical system evolves
from a state governed by the asymptotic Hamiltonian
to the physical state described by the total Hamiltonian.
In the two-potential
formalism of that,  an intermediate Hamiltonian
must be defined and
connection made first to the
asymptotic Hamiltonian
and then to the total Hamiltonian. The two-potential picture is a direct
consequence of Kato's chain rule for wave operators \cite{kato}, which holds
even for Coulomb interactions \cite{bencze}. 
In a ``three-potential'' picture~\cite{pzsc} a three-body Coulomb
scattering process can be viewed as three consecutive scattering
processes by which  the asymptotic channel Hamiltonian
connects to the total one via  two intermediate Hamiltonians. 
This process can be formulated in terms of
a set of the Faddeev and Lippmann--Schwinger
integral equations. Numerically, these integral equations
can be solved by using a Coulomb--Sturmian space representation.
Hereafter I designate both the representation and the method
by the label CS. The (CS) method has been used before and
it has worked very well
for bound-state problems with repulsive~\cite{pzwp} and with
attractive~\cite{pzatom} Coulomb interactions.
It has also been used to analyze
$p-d$ scattering at energies below breakup~\cite{pzsc}
and the results
agree very well with those obtained from other 
calculations \cite{chen}.

In this article, I seek
 to extend the scope of this novel (CS) method
to energies above the breakup threshold. I shall show that the 
Faddeev and Lippmann--Schwinger integral equations deduced from a
``three-potential'' picture by applying Kato's chain rule 
possess compact kernels and so have
unique solutions to the three-body Coulomb problem
for all energies. I will also show how one can 
calculate the Green's operator which
contains all the asymptotically relevant terms
for arbitrary complex energies.

I assume that the subsystem interaction, $v_\alpha$, 
of two elements in a three-body system
is a Coulomb-like one, which is then split into
 short-range and  long-range terms as
\begin{equation}
v_\alpha(\xi_\alpha)=v_\alpha^{(s)}(\xi_\alpha)+
v_\alpha^{(l)}(\xi_\alpha).
\label{pot}
\end{equation}
Here $\alpha$ denotes the subsystem and superscripts
$s$ and $l$ indicate the short- and long-range
attributes, respectively,
with ${\xi}_\alpha$ and ${\eta}_\alpha$ being
the usual configuration space Jacobi coordinates.
The splitting should be performed in such a way that $v_\alpha^{(l)}$
does not support any bound states.
The total Hamiltonian is denoted by $H$ and the 
asymptotic channel Hamiltonian is defined by
\begin{equation}
H_\alpha = 
h^0_{\xi_\alpha}+h^0_{\eta_\alpha} + v_\alpha
= H^0 + v_\alpha, 
\label{halpha0}
\end{equation}
where $h^0$ and $H^0$ are the two-body and the three-body 
kinetic energy operators respectively.
 The asymptotic state
$| {\Phi}_{\alpha}\rangle$ is 
an eigenstate of $H_\alpha$, i.e.
\begin{equation}
H_\alpha | {\Phi}_{\alpha}\rangle = E_\alpha | {\Phi}_{\alpha}\rangle,
\end{equation}
where  $\langle \xi_\alpha \eta_\alpha |
{\Phi}_{\alpha}\rangle =
\langle \eta_\alpha|
{\chi}_{\alpha}\rangle \langle
\xi_\alpha | {\phi}_{\alpha} \rangle$
is a product of a continuum state in coordinate $\eta_\alpha$ and a 
bound or continuum state in the two-body subsystem $\xi_\alpha$.
If ${\phi}_{\alpha}$ is a bound
eigenstate,  asymptotically there is a two-fragment channel.
If it is a scattering
state, the asymptote is a three-fragment channel. 
The scattering state, $\Psi_\alpha$, which evolves from the
asymptotic state, is given by applying the 
 $\Omega^{(\pm)}(H,H_\alpha)$ M{\o}ller operators onto the asymptotic state
\begin{equation}
| {\Psi}_{\alpha}^{(\pm)}\rangle
= \Omega^{(\pm)}{(H ,H_\alpha)} | {\Phi}_{\alpha}\rangle.
\label{Psi}
\end{equation}
The  M{\o}ller operators are defined by the limit
\begin{equation}
\Omega^{(\pm)}{(H ,H_\alpha)} = \mathop{\rm s-lim}_{t \rightarrow \mp\infty}
  \exp{[{\mathrm{i}}Ht]} \exp{[-{\mathrm{i}}H_\alpha(t)]},
 \label{moller1}
\end{equation}
where $H_\alpha(t)=H_\alpha t + A(t)$ with $A(t)$ being the Dollard's
distortion operator \cite{dollard}.

In the spirit of Kato's chain rule, I  introduce two intermediate
Hamiltonians, the 
channel long-range Hamiltonian
\begin{equation}
H^{(l)}_\alpha=H_\alpha +  v_\beta^{(l)} +v_\gamma^{(l)},
\label{hca}
\end{equation}
and the channel distorted long-range Hamiltonian
\begin{equation}
\widetilde{H}_\alpha=H_\alpha + u_\alpha^{(l)}.
\label{halpha}
\end{equation}
The auxiliary potential
$u_\alpha^{(l)}=u_\alpha^{(l)}(\eta_\alpha)$ is defined  such 
that it does not
support any bound states and  has the asymptotic form
$u_\alpha^{(l)} \sim {e_\alpha (e_\beta+e_\gamma) }/{\eta_\alpha}$
as ${\eta_\alpha \to \infty}$. In fact, $u_\alpha^{(l)}$
is an effective Coulomb interaction between the center of 
mass of the subsystem $\alpha$ (with
charge $e_\beta+e_\gamma$) and the third particle
(with charge $e_\alpha$).
Now, according to Kato's chain rule the M{\o}ller operator
(\ref{moller1}) can be written in the form
\begin{eqnarray}
\Omega^{(\pm)}{(H, H_\alpha)} &=& \Omega^{(\pm)}{(H, H_\alpha^{(l)})} 
\nonumber \\
& & \times \Omega^{(\pm)}{(H_\alpha^{(l)}, \widetilde{H}_\alpha)}
\Omega^{(\pm)}{ (\widetilde{H}_\alpha, H_\alpha)}.
\label{3moller}
\end{eqnarray}

The last term, $\Omega^{(\pm)}{ (\widetilde{H}_\alpha, H_\alpha)}$,
which essentially describes a two-body scattering in the Coulomb-like
potential $u_\alpha(\eta_\alpha)$,
exists in Dollard's sense, and by applying it to the 
channel state  gives
\begin{equation}
| \widetilde{\Phi}_{\alpha}^{(\pm)} \rangle = 
\Omega^{(\pm)}{ (\widetilde{H}_\alpha, H_\alpha)} | 
\Phi_{\alpha} \rangle=
| \widetilde{\chi}^{(\pm)}_{\alpha}\rangle 
| {\phi}_{\alpha} \rangle
\end{equation}
with $\langle \eta_\alpha |
\widetilde{\chi}^{(\pm)}_{\alpha}\rangle$ being a scattering solution in
the Coulomb-like potential $u_\alpha^{(l)}$. 

Now I will show that the middle term in (\ref{3moller}) exists in the ordinary
sense \cite{taylor}
\begin{equation}
\Omega^{(\pm)} {(H_\alpha^{(l)} ,\widetilde{H}_\alpha)} =
 \mathop{\rm s-lim}_{t \rightarrow \mp\infty}
  \exp{[{\mathrm{i}} H_\alpha^{(l)} t]}
  \exp{[-{\mathrm{i}}\widetilde{H}_\alpha t]}.
 \label{moller2}
\end{equation}
The necessary condition for the existence of the strong limit in
(\ref{moller2})  is that  the potential
\begin{equation}
U^\alpha=H_\alpha^{(l)}-\widetilde{H}_\alpha  = 
v_\beta^{(l)}+v_\gamma^{(l)}- u_\alpha^{(l)}
 \label{U}
\end{equation}
should decrease in configuration space more rapidly than 
does the Coulomb potential \cite{amrein}.
Indeed, the auxiliary potential $u_\alpha^{(l)}$ has been
 constructed so that it cancels the
long-range potentials $v_\beta^{(l)}+v_\gamma^{(l)}$ in $U^\alpha$
for the two-fragment channel $\alpha$ 
in which the particle labeled with 
$\alpha$ is in continuum state, while
the particles labeled by $\beta$ and $\gamma$
 form a bound state.
Also, in the three-fragment channel of $H_\alpha$, 
where the particles labeled by $\beta$ and $\gamma$ 
are in continuum state,
$U^\alpha$ decreases with radius faster than 
does the Coulomb potential.
The term $U^\alpha$ would behave like a Coulomb potential
in the two-fragment channels,
$\beta$ and $\gamma$, where particles $\alpha$ and $\gamma$ or 
$\alpha$ and $\beta$ form bound clusters, but for $H_\alpha^{(l)}$
there are no two-fragment channels $\beta$ and $\gamma$. 

Applying $\Omega^{(\pm)} {(H_\alpha^{(l)} ,\widetilde{H}_\alpha)}$ to the
``free'' state $ |\widetilde{\Phi}_{\alpha}^{(\pm)} \rangle$ yields
\begin{equation}
| \Phi_{\alpha}^{(l)(\pm)} \rangle=
\Omega^{(\pm)} {(H_\alpha^{(l)} ,\widetilde{H}_\alpha)} 
|\widetilde{\Phi}_{\alpha}^{(\pm)} \rangle.
\end{equation}
Sandhas has shown in Ref.\ \cite{sandhas}, that if for a three-body
system only one of the potentials supports bound states,  no 
rearrangement channels are possible, 
and therefore a single Lippmann--Schwinger
equation is completely sufficient to guarantee the uniqueness of the
solution. The Hamiltonian $H_\alpha^{(l)}$ possesses this property, thus
the Lippmann--Schwinger integral equation which is derived 
from (\ref{moller2}), 
\begin{equation}
| \Phi_{\alpha}^{(l)(\pm)} \rangle = |
\widetilde{\Phi}_{\alpha}^{(\pm)} \rangle +
\widetilde{G}_\alpha (E_\alpha \pm \mathrm{i} 0 )
 U^\alpha | \Phi_{\alpha}^{(l)(\pm)} \rangle,
\label{lswf}
\end{equation}
with $\widetilde{G}_\alpha (z)=(z-\widetilde{H}_\alpha)^{-1}$,
provides unique solution for this auxiliary three-body system.
Additionally, Eq.~({\ref{lswf}) possesses a compact kernel
because in the asymptotically accessible
region, channel $\alpha$, $\widetilde{G}_\alpha$
is linked solely to the asymptotic behavior of
 $| \Phi_{\alpha}^{(l)(\pm)} \rangle$,
and the source term, $U^\alpha$,
is of shorter range than the Coulomb interaction.
Thus, in this particular Lippmann-Schwinger equation
$U^\alpha$ can be approximated by finite rank terms.
Note, that a similar equation holds for $G_\alpha^{(l)}=
(z-H_\alpha^{(l)})^{-1}$:
\begin{equation}
G_\alpha^{(l)}=\widetilde{G}_\alpha+\widetilde{G}_\alpha U^\alpha
G_\alpha^{(l)}.
\label{lsgc}
\end{equation}

Finally, $\Omega^{(\pm)}{(H, H_\alpha^{(l)})}$, which also exists
in the usual sense, leads to
the Faddeev--Noble integral equations \cite{noble},
\begin{equation}
|\psi_{\beta}^{(\pm)} \rangle=\delta_{\beta \alpha}
|\Phi_{\alpha}^{(l)(\pm)}\rangle + G_\beta^{(l)} 
(E_\alpha \pm {\mathrm{i}} 0)
v^{(s)}_\beta \sum_{\gamma\neq\beta}
|\psi_{\gamma}^{(\pm)}\rangle,
\label{fn-eq}
\end{equation}
where 
$\alpha, \beta, \gamma$ form a cyclic permutation.
Merkuriev has shown that the kernels of these equations are compact for all
energies \cite{merkuriev}, and thus 
the potential operators can be approximated well by finite rank terms.
So, the set of Lippmann--Schwinger 
and Faddeev--Noble integral equations derived from a
``three-potential'' picture for the three-body Coulomb scattering problem,
are uniquely solvable and possess compact kernels for all energies.

In this derivation a crucial point is that
one Lippmann--Schwinger equation of the type given in
Eq.~(\ref{lswf}) suffices for a unique solution
to be found. This is the case if one
imposes a condition on $v^{(l)}$ that it should not
support bound states. 
This condition is satisfied if
all the Coulomb interactions in the three-body 
system are repulsive. With a three-nucleon system then,
one would take
 $v_\alpha^{(s)}$ as the nuclear potential
and $v_\alpha^{(l)}$ as the Coulomb interaction.
The situation is more complicated if 
some of the Coulomb interactions
are attractive. There are an  infinite number of bound states
associated with  
 an attractive Coulomb potential; a fact closely related to
its long-range character. 
Nevertheless, the procedure is applicable to such systems
whenever the energy is below
the three-body breakup threshold.
For such cases
one Lippmann--Schwinger equation is still sufficient
to find solutions.

The ``three-potential'' integral equations have been solved 
previously by using a CS representation~\cite{pzsc,pzwp}.
With $n$ and $l$ being the radial and
orbital angular momentum quantum numbers respectively,
the CS functions $|nl\rangle$
form a biorthonormal
discrete basis in the two-body Hilbert space; the biorthogonal
partner defined  by $\langle r |\widetilde{nl}\rangle= 
r^{-1} \langle r |{nl}\rangle$.
Since the three-body Hilbert space is a
direct sum of two-body
Hilbert spaces, an appropriate basis in angular momentum
representation is
 the direct product
\begin{equation}
| n \nu l \lambda \rangle_\alpha =
[ | n l \rangle_\alpha \otimes | \nu
\lambda \rangle_\alpha ] , \ \ \ \ (n,\nu=0,1,2,\ldots).
\label{cs3}
\end{equation}
Here $l$ and $\lambda$ denote the angular momenta associated 
with  the coordinates $\xi$ and $\eta$, respectively. 
In this basis the completeness relation
takes the form 
\begin{equation}
{\bf 1} =\lim\limits_{N\to\infty} \sum_{n,\nu=0}^N |
 \widetilde{n \nu l
\lambda} \rangle_\alpha \;\mbox{}_\alpha\langle
{n \nu l \lambda} | =
\lim\limits_{N\to\infty} {\bf 1}_{N}^\alpha;
\end{equation}
a sum over the angular momenta being assumed implicitly.
Note that in the three-body Hilbert space, 
three equivalent bases belonging to fragmentation
$\alpha$, $\beta$ and $\gamma$ are possible.

To proceed, it is convenient to  approximate Eqs.~(\ref{fn-eq})  by
\begin{equation}
|\psi_{\beta}^{(\pm)} \rangle=\delta_{\beta \alpha}
|\Phi_{\alpha}^{(l)(\pm)}\rangle + G_\beta^{(l)}
{\bf 1}_{N}^\beta
v^{(s)}_\beta \sum_{\gamma\neq\beta}  {\bf 1}_{N}^\gamma
|\psi_{\gamma}^{(\pm)}\rangle,
\label{feqsapp}
\end{equation}
\noindent
with the short-range potential
$v_\alpha^{s}$ in the three-body
Hilbert space taken to have a separable form, viz.
\begin{equation}
v_\alpha^{s}\approx \sum_{n,\nu ,n^{\prime },
\nu ^{\prime }=0}^N|\widetilde{n\nu l\lambda }\rangle _\alpha \;
\underline{v}_{\alpha \beta }^{s}
\;\mbox{}_\beta \langle \widetilde{n^{\prime }
\nu ^{\prime }l^{\prime }\lambda
^{\prime }}|,  \label{sepfe}
\end{equation}
where $\underline{v}_{\alpha \beta}^{s}=
\mbox{}_\alpha \langle n\nu l\lambda |
v_\alpha^{s}|n^{\prime }\nu ^{\prime
}l^{\prime }{\lambda }^{\prime }\rangle_\beta$.
In Eq.~(\ref{sepfe}) the ket and bra states are defined
for different fragmentations, depending on the
environment of the potential operators in the equations.
A similar approximation is made on the potential $U^\alpha$
in Eqs.~(\ref{lsgc}) and (\ref{lswf}),
with bases of the same fragmentation $\alpha$ 
applied  on both sides of the operator.
Thus,  by truncating  the short-range operators in Eqs.~(\ref{fn-eq}),
(\ref{lsgc}) and (\ref{lswf}), the set of linear integral
equations reduces to an analogous set of linear algebraic 
equations with the operators replaced by their matrix representations.
Calculation of $\underline{v}_{\alpha \beta}^{s}$ and of
$\underline{U}^\alpha$ can be made as shown previously~\cite{pzwp}.
A similar procedure enables
calculation of the matrix
elements
$\underline{\widetilde{G}}_{\alpha}=
\mbox{}_\alpha \langle \widetilde{n\nu l\lambda} |
\widetilde{G}_\alpha |  \widetilde{ n^{\prime }\nu ^{\prime
}l^{\prime }{\lambda }^{\prime }}\rangle_\alpha $
 with three-body bound states~\cite{pzwp}.
The Green's operator $\widetilde{G}_\alpha$
is a resolvent of the sum of two commuting Hamiltonians,
$h_{\xi _\alpha }=h^0_{\xi _\alpha }+v_\alpha$ and 
$h_{\eta _\alpha }= h^0_{\eta_\alpha} + u_\alpha^{(l)}$,
which act in different two-body Hilbert spaces.
Thus, using  the convolution theorem~\cite{bianchi},
which is a direct consequence of Dunford's operator 
calculus~\cite{dunford}, the three-body Green's operator 
$\widetilde{G}_\alpha$ equates to 
a convolution integral of two-body Green's operators, i.e.
\begin{eqnarray}
\widetilde{G}_\alpha (z)&=&(z-h_{\xi _\alpha }-
h_{\eta _\alpha })^{-1} \nonumber \\
&=& \frac 1{2\pi \mathrm{i}}\oint_C
dz^\prime \,g_{\xi_\alpha }(z-z^\prime)\;
g_{\eta_\alpha}(z^\prime),
 \label{contourint}
\end{eqnarray}
where
$g_{\xi_\alpha}(z)=(z-h_{\xi_\alpha})^{-1}$  and
$g_{\eta_\alpha}(z)=(z-h_{\eta_\alpha})^{-1}$.
The contour $C$ should be taken  counterclockwise
 around the continuous spectrum of $h_{\eta_\alpha }$
so that
$g_{\xi_\alpha }$ is analytic in the domain encircled
by $C$. For bound-state energies the spectra of the two Green's
operators are well separated and this condition can be fulfilled 
easily~\cite{pzwp}. Also, below breakup threshold,
where the bound-state pole of $g_{\xi_\alpha}$ meets the
continuous spectrum of $g_{\eta_\alpha}$,
contour integration can still be performed~\cite{pzsc}.
 For other positive energy scattering problems, however,
the continua overlap so the applied contours are not
viable.

But there exists a contour which is valid for all $z$ of physical
interest.  Besides positive real values of $z$
that arise with scattering above the breakup threshold,
this contour can deal with complex values of $z$
having negative imaginary parts 
that are needed for resonant-state
calculations. In this approach first 
one must shift the spectrum of $g_{\xi_\alpha }$ by
taking  $z=E +{\mathrm{i}}\varepsilon$  with
positive $\varepsilon$. By doing so, 
the two spectra are well separated and 
the spectrum of $g_{\eta_\alpha }$ can be encircled.
Next the contour $C$ is deformed analytically
in such a way that the upper part descends to the second
Riemann sheet of $g_{\eta_\alpha}$, while
the lower part of $C$ can be detoured away from the cut
 [see  Fig.~\ref{fig1}]. The contour still
encircles the branch cut singularity of $g_{\eta_\alpha}$,
but in the  $\varepsilon\to 0$ limit it now
avoids the singularities of $g_{\xi_\alpha}$.
 Moreover, by continuing
to negative values of  $\varepsilon$,
the branch cut and pole
singularities of $g_{\xi_\alpha}$ move onto
 the second Riemann sheet
of $g_{\eta_\alpha}$ and, at the same time, the  branch cut
of $g_{\eta_\alpha}$ moves onto the second Riemann sheet
of $g_{\xi_\alpha}$. Thus, the mathematical conditions for
the contour integral representation
of $\widetilde{G}_\alpha (z)$ in 
Eq.~(\ref{contourint}) can be fulfilled for all energies.
In this respect there is only a gradual difference between the
bound- and resonant-state calculations and scattering ones
below and above the breakup threshold.

The matrix elements $\underline{\widetilde{G}}_\alpha$
can be cast in the form
\begin{eqnarray}
{\underline{\widetilde{G}}_\alpha (z) }
& =&\frac{1}{2 \pi \mathrm{i}} \oint_C dz^\prime  \
\mbox{}_\alpha\langle
 \widetilde{ n l}|
g_{\xi_\alpha}(z-z^\prime) |
\widetilde{ n^{\prime}{l^{\prime}}}
\rangle_\alpha\nonumber \\
&& \ \times \
\mbox{}_\alpha\langle
\widetilde{ \nu \lambda }|
g_{\eta_\alpha}(z^\prime) | \widetilde{
\nu^{\prime}{\lambda^{\prime}} }
\rangle_\alpha\ .
\label{contourint2}
\end{eqnarray}
The matrix elements of the two-body Green's 
functions in
the integrand are known for all complex energies
from two-body state properties~\cite{cpc,jmp}.
In this formalism the Faddeev components appear
 as linear combinations
of functions
\begin{eqnarray}
\lefteqn{ \langle \xi_\alpha \eta_\alpha |\psi _\alpha \rangle \sim
\langle \xi_\alpha \eta_\alpha |
{{\widetilde{G}}_\alpha (z) }| \widetilde{ n \nu l \lambda}
\rangle_\alpha = }\nonumber \\
& &\frac{1}{2 \pi \mathrm{i}} \oint_C dz^\prime  \
\langle \xi_\alpha | g_{\xi_\alpha}(z-z^\prime) |
\widetilde{ n l}
\rangle_\alpha \
\langle \eta_\alpha | g_{\eta_\alpha}(z^\prime) | \widetilde{
\nu \lambda }
\rangle_\alpha,  \label{contourwf}
\end{eqnarray}
which are convolution integrals of Coulomb-like functions \cite{cpc}.
 
In this paper I have shown that 
the set of Faddeev and Lippmann--Schwinger equations
for the three-body Coulomb problem 
derived from the ``three-potential''
picture~\cite{pzsc} possess compact kernels for all energies.
I have found an analytic representation of the channel-distorted
Coulomb Green's operators  in terms of a convolution
integral of two-body Green's operators.
The method facilitates the solution of
integral equations in which the exact bound- scattering- or resonant-state
Coulomb asymptotics are automatically incorporated.
Solution of  these equations in a Coulomb--Sturmian space
 representation is most practical,
as then there is an analytic representation
of the terms in the contour integrals.

The author is most grateful for the useful discussions
he has had with Profs. W. Sandhas (Bonn), H. V. von Geramb (Hamburg)
and K. A. Amos (Melbourne).
This work has been supported by OTKA under Contracts No.\ T17298
and No.\ T026233.

\begin{figure}
\psfig{file=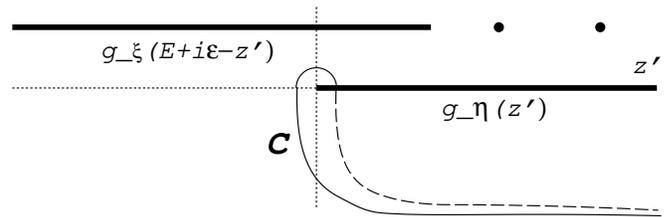,width=8.7cm}
\caption{Analytic structure of $g_{\xi_\alpha }(z-z^\prime)\;
g_{\eta_\alpha}(z^\prime)$ as a function of $z^\prime$ with
$z=E+{\mathrm{i}}\varepsilon$, $E>0$, $\varepsilon>0$.
The contour $C$ encircles the continuous spectrum of 
$h_{\eta_\alpha}$. A part of it, which goes on the second
Riemann-sheet of $h_{\eta_\alpha}$, is drawn by broken line.}
\label{fig1}
\end{figure}

\end{document}